\documentclass[12pt]{article}
\pdfoutput=1
\usepackage{amsmath,amssymb}
\usepackage{graphicx}
\usepackage{subfigure}
\usepackage[pdftex]{hyperref}
\DeclareGraphicsExtensions{.eps,.bmp,.wmf,.jpg,.pdf}
\numberwithin{equation}{section}
\def\be{\begin{equation}}
\def\ee{\end{equation}}

\def\bea{\begin{eqnarray}}
\def\eea{\end{eqnarray}}

\title{The holographic principle and the cosmological constant}
\author{L.N. Granda\thanks{ngranda@univalle.edu.co, ngranda@um.es}\\{\small\it Departamento de Fisica, Universidad del Valle, A.A. 25360 Cali, Colombia}\\
{\small\it Departamento de Fisica, Universidad de Murcia, 30100 Murcia, Spain}}

\date{}
\begin{document}
\maketitle

\begin{abstract}
\noindent We propose an infrared cut-off for the holographic density which incorporates among others a constant term, that produces the effect of the cosmological constant, improving the results of previously considered holographic models based on local quantities. The inclusion of constant term is interpreted as a natural first approximation for the infrared cutoff which is associated with the vacuum energy, and the additional terms guarantee an appropriate evolutionary scenario that fits the astrophysical observations. Cosmological constraints on the model have been studied using the observational data.
\end{abstract}

\section{Introduction}
\noindent 
The astrophysical data from distant Ia supernovae observations \cite{riess}, \cite{riess1}, \cite{perlmutter}, \cite{perlmutter1}, \cite{astier}, \cite{kowalski}, \cite{amanullah}, \cite{hicken}, cosmic microwave background anisotropy \cite{spergel}, \cite{spergel1}, \cite{komatsu}, \cite{larson}, and large scale galaxy surveys \cite{tegmark}, \cite{tegmark1}, \cite{abazajian}, \cite{abazajian1}, \cite{tegmark2}, \cite{einsenstein}, \cite{tegmark3}, all indicate that the current Universe is not only expanding, it is accelerating due to some kind of  negative-pressure form of matter known as dark energy (\cite{peebles}, \cite{copeland},\cite{turner},\cite{sahni}, \cite{sahnii}, \cite{padmanabhan}). The combined analysis of cosmological observations also suggests that the universe is spatially flat, and  consists of about $\sim 1/3$
of dark matter (the known baryonic and nonbaryonic dark matter), distributed in clustered structures (galaxies, clusters of galaxies, etc.) and $\sim 2/3$ of homogeneously distributed (unclustered) dark energy with negative pressure. Despite the high percentage of the dark energy component, its nature as well as its cosmological origin remain unknown at present and a wide variety of models have been proposed to explain the nature of the dark energy and the accelerated expansion (see \cite{peebles}-\cite{padmanabhan} for review).
Among the different models of dark energy, the holographic dark energy approach is quite interesting as it incorporates some concepts of the quantum gravity known as the holographic principle (\cite{bekenstein, thooft, bousso, cohen, susskind}),which first appeared in the context of black holes \cite{thooft} and later extended by Susskind \cite{susskind} to string theory. According to the holographic principle, the entropy of a system scales not with its volume, but with its surface area. In the cosmological context, the holographic principle will set an upper bound on the entropy of the universe \cite{fischler}. In the work \cite{cohen}, it was suggested that in quantum field theory a short distance cut-off $\Lambda$ is related to a long distance cut-off (infra-red cut-off $L$) due to the limit set by black hole formation, namely, if is the quantum zero-point energy density caused by a short distance (UV) cut-off, the total energy in a region of size $L$ should not exceed the mass of a black hole of the same size, thus $L^3\Lambda^4\leq LM_p^2$. Applied to the dark energy issue, if we take the whole universe into account, then the vacuum energy related to this holographic principle is viewed as dark energy, usually called holographic dark energy \cite{cohen} \cite{hsu}, \cite{li}. The largest $L$ allowed is the one saturating this inequality so that the holographic dark energy density is defined by the equality $\rho_\Lambda=3c^2M_p^2L^{-2}$, where $c^2$ is a numerical constant and $M_p^{-2}=8\pi G$.\\
As is well known, the Hubble horizon $H^{-1}$ as the infrared cut-off, gives an equation of state parameter (EoS) equal to zero, behaving as pressureless matter which cannot give accelerated expansion, and the particle horizon gives an EoS parameter larger than $-1/3$, which is not enough to satisfy the current observational data. The infrared cut-off given by the future event horizon yields the desired result of accelerated expansion with an EoS parameter less than $-1/3$, despite the fact that it is based on non local quantities and has problems with the causality \cite{li}. An holographic dark energy model which is based on local and non local quantities have been considered in \cite{sergei}, \cite{sergei1}. Based mainly on dimensional arguments, in \cite{granda}, \cite{granda2} an infrared cut-off for the holographic density was proposed, which describes the dark energy in good agreement with the astrophysical data, and may explain the cosmic coincidence.
This model exhibits quintom nature (for some values of the parameters) without the need to introduce any exotic matter, and has also proven to be useful in the reconstruction of different scalar field models of dark energy which reproduce the late time cosmological dynamics in a way consistent with the observations \cite{granda3,granda4}.\\ 
Despite the available proposals, the theoretical root of the holographic dark energy is still unknown mostly because our ignorance about the microscopic nature of quantum gravity and the origin of dark energy. This translates into a lack of details about the holographic cut-off, and therefore in the present paper we assume that this cut-off should be encoded in a function that depends on natural cosmological quantities like the Hubble parameter, its time derivative and also should contain a constant term, which represents the first approximation to vacuum energy density (which could be interpreted, in the context of the holographic principle as giving quantum gravity nature to the cosmological constant). 
The constant term would act as the first natural approximation to the infrared cut-off for the holographic density, since this term may be identified with a constant vacuum energy density, that adjusts very well to the current observational data. A discussion about the cosmological constant and the holographic principle is given in \cite{horava}.
We propose an holographic density given by a general function of the form $f(H,\dot{H})$, where $f$ should be expanded in powers of his arguments, based on their smallness (compared to the squared Planck mass $M_p^2$). The coefficients of the different terms of this expansion must be found by fitting the model with current observations.
To obtain the cosmological constraints on the model, we will use the SNIa Constitution sample (\cite{hicken}), complemented with the CMB (cosmic micro wave background) anisotropy and BAO (baryon acoustic oscillation) data. In section 2 we review the main aspects of the model, in section 3 we include the matter content and fit the parameters using the joint Constitution, CMB and BAO data analysis. In section 4 we give a discussion.
\section{The Model}
The main inequality discussed above $L^3\Lambda^4 \leq LM_p^2$, leads to the restriction involving the UV and IR cut-offs $\Lambda^4\leq M_p^2 L^{-2}$.  Saturating this inequality we come to the concept of holographic dark energy $\rho_{\Lambda}$. To define the infrared cut-off and hence the holographic dark energy density, we propose for $L^{-2}$ a general function which should depend on the main cosmological parameters $H$ and $\dot{H}$, and is measured in units of $(length)^{-2}$, defining the holographic density as follows (from now on we will set $8\pi G=M_p^{-2}=1$)
\begin{equation}\label{eq1}
\rho_\Lambda=3f(H^2,\dot{H})
\ee
where $H=\dot{a}/a$ is the Hubble parameter, and the function $f$ can be Taylor expanded in powers of his arguments provided $H^2$ and $\dot{H}$ are at the present very small quantities measured in Planck mass units, giving
\be\label{eq3}
f(H^2,\dot{H})=\lambda+\alpha H^2+\beta \dot{H}
\ee
where we neglected higher order powers of $H^2$ and $\dot{H}$, and $\lambda$, $\alpha$ and $\beta$ are constants to be determined. This model generalizes the one presented in (\cite{granda},\cite{granda2}). The novelty of the present proposal takes root in the presence of the constant term $\lambda$ which we consider as the first natural approximation to the infrared cut-off ($L\sim \lambda^{-1/2}$), and should be associated with the contribution of the constant energy density of the vacuum. As we shall see bellow, the cosmological dynamics depends on $\alpha$ and $\beta$ only through the combination $(\alpha-1)/\beta$ (i.e. cosmological observations fix only this combination) leaving one free parameter. To fix the free parameter we can use geometrical criteria in order to reduce the last two terms in (\ref{eq3}) to a term proportional to Ricci scalar, but the dynamics is independent of the election of this free parameter. \\
\noindent To analyze some cosmological consequences of the model (\ref{eq3}), we first consider the case of an universe dominated by dark energy. The Friedmann equation in absence of matter is
\begin{equation}\label{eq5}
H^2=\frac{1}{3}\rho_\Lambda=\lambda+\alpha H^2+\beta \dot{H}
\end{equation}
\noindent Setting $x=\ln{a}$, we can rewrite (\ref{eq5}) as follows
\begin{equation}\label{eq6}
H^2=\lambda+\alpha H^2+\frac{\beta}{2}\frac{dH^2}{dx}
\end{equation}
\noindent Introducing the scaled Hubble expansion rate $\tilde{H}=H/H_0$, where $H_0$ is the present value of the Hubble parameter (for $x=0$), the above Friedman equation becomes
\begin{equation}\label{eq7}
\tilde{H}^2=\Omega_{\Lambda}+\alpha\tilde{H}^2
+\frac{\beta}{2}\frac{d\tilde{H}^2}{dx}
\end{equation}
where $\Omega_{\Lambda}=\lambda/H_0^2$ Solving the equation (\ref{eq7}), we obtain
\begin{equation}\label{eq8}
\tilde{H}^2=\Omega_{\Lambda}+\Omega e^{-2x(\alpha-1)/\beta}=\Omega_{\Lambda}+\Omega (1+z)^{2(\alpha-1)/\beta}
\end{equation}
where $\Omega_{\Lambda}=\Omega_{\lambda}/(1-\alpha)$, $\Omega$ is an integration constant and the last equality is written in the redshift variable (by using $e^{-x}=1+z$). 
Note that even without the matter component, this solution is interesting enough as it contains the cosmological constant, and the second term may give matter-like behavior if the constants $\alpha$ and $\beta$ satisfy the restriction $(\alpha-1)/\beta=3/2$, giving rise to the $\Lambda$CDM model. Evaluating the Eq. (\ref{eq8}) at $x=0$ (the flatness condition), it follows
\be\label{eq9}
\Omega_{\Lambda}+\Omega=1.
\ee
Thus $\Omega_{\Lambda}$ represents the constant component of vacuum energy density and $\Omega$ stands for the new form of ``matter'', both quantities coming from the holographic principle.  Additional constraints may be used to find appropriate values for the parameters. These constraints are related with the present observational value of the dark energy EoS, which between the margins of errors can be set to $\text{w}\approx-1.1$, and the transition deceleration-acceleration, which occurs in the region $z_T\sim 0.5-0.7$ ($z_T$ is the transition redshift). The EoS in terms of the redshift is given by
\be\label{eq10}
\text{w}(z)=-1+\frac{1}{3}\frac{(1+z)}{\tilde{H}^2}\frac{d(\tilde{H}^2)}{dz}
\ee
replacing $\tilde{H}^2$ from (\ref{eq8}) it follows that 
\be\label{eq11}
\text{w}(z)=-1-\frac{2(1-\alpha)(1-\Omega_{\Lambda})(1+z)^{2(\alpha-1)/\beta}}{3\beta\left[\Omega_{\Lambda}+(1-\Omega_{\Lambda})(1+z)^{2(\alpha-1)/\beta}\right]}
\ee
from Eq. (\ref{eq8}) the deceleration parameter $q=-a\ddot{a}/\dot{a}^2$, is given by
\be\label{eq12}
q(z)=-1-\frac{(1-\alpha)(1-\Omega_{\Lambda})(1+z)^{2(\alpha-1)/\beta}}{\beta\left[\Omega_{\Lambda}+(1-\Omega_{\Lambda})(1+z)^{2(\alpha-1)/\beta}\right]}
\ee
Note that the Hubble parameter (\ref{eq8}) and the Eqs. (\ref{eq11}) and (\ref{eq12}) depend on $\alpha$ and $\beta$ only through the combination $(\alpha-1)/\beta$.
Therefore, for a given $\text{w}(z=0)=\text{w}_0$ and $q(z_T)=0$, we can find the combination $(\alpha-1)/\beta$ and $\Omega_{\Lambda}$. Note also that $\Omega_{\Lambda}=1$ 
gives $\text{w}=-1$, as expected from a universe dominated by the cosmological constant. 
It is worth examining the special case of $\alpha=1$. Integrating the Eq. (\ref{eq7}) for $\alpha=1$, and using (\ref{eq10}) to the define the corresponding EoS, gives
\be\label{eq12a}
\text{w}(z)=-1+\frac{2\Omega_{\lambda}}{3C\beta+6\Omega_{\lambda}\log(1+z)},
\ee
where $C$ is the integration constant. At $z=0$ gives $\text{w}_0=-1+2\Omega_{\lambda}/(3C\beta)$. Thus depending on the relative sign of $\beta$ and $C$, we can have quintessence or phantom behavior. An interesting limit of Eq. (\ref{eq7}) can be obtained when $\dot{H}<<H^2$, which is equivalent to one of the slow-roll conditions used in inflationary models. In this limit $H^2=\lambda/(1-\alpha)$, which gives rise to exponential expansion ($\alpha\neq 1$).
%%%%%%%%%%%%%%%%%%%%%%%%%%%%%%%%%%%%%%%%%%%%%%%%%%%%%%%%%%%%%%%%%%%%%%%%%%%%%%%%%%%%%%%%%%%%%%%%%%%%%%%%%%%%%%%%%%%%%%%%%%%%%%%%%%%%%%%%%%%%%%%%%%%%%%%%%%%%%%%%%%%%%%%%
\section{Holographic density with matter}
In order to construct a more realistic scenario, we include the contribution of matter, and by using the observational data we can define the parameters of the model.
Adding the contribution of matter, the Friedmann equation becomes
\be\label{eq13}
H^2=\frac{1}{3}\rho_{m0} e^{-3x}+\lambda+\alpha H^2+\frac{\beta}{2}\frac{dH^2}{dx},
\ee
with the solution given by (in terms of the scaled Hubble parameter)
\be\label{eq14}
\tilde{H}^2=\Omega_{m0} e^{-3x}+\Omega_{\Lambda}+\Omega e^{-2x(\alpha-1)/\beta}+\frac{(2\alpha-3\beta)\Omega_{m0}}{3\beta-2\alpha+2} e^{-3x}
\ee
where $\Omega_{m0}=\rho_{m0}/(3H_0^2)$, $\Omega$ is the integration constant and the last three terms represent the contribution of the holographic dark energy. The third term provides the possibility of crossing the phantom barrier, giving rise to quintom behavior. Note that the last term in (\ref{eq14}) makes contribution to the matter sector, which 
indicates that the present holographic density also contributes to the presureless matter component of the universe, which is about 27\% of the total.
In what follows we will consider the sum of the terms proportional to $e^{-3x}$ as the matter component (in fact is an effective matter component) and the Eq. (\ref{eq14}) will be written as 
\be\label{eq15}
\tilde{H}^2=\Omega'_{m0} e^{-3x}+\Omega_{\Lambda}+\Omega e^{-2x(\alpha-1)/\beta},
\ee
where
\be\label{eq16}
\Omega_{m0}'=\frac{2}{3\beta-2\alpha+2}\Omega_{m0}
\ee
and $\Omega_{m0}'$ should be adjusted with the data. According to current observations, the main contribution to dark energy comes from the constant term in Eq. (\ref{eq15}), and the third term may be responsible for possible quintom behavior. The coefficients in Eq. (\ref{eq15}) are subject to the flatness condition (at $x=0$)
\be\label{eq17} 
\Omega'_{m0}+\Omega_{\Lambda}+\Omega=1
\ee
It is important to note here that the net contribution to the present density parameter of the dark energy comes from the sum of the last two terms in Eq. (\ref{eq17}), which we can resume as
\be\label{eq18}
\Omega_{DE}=\Omega_{\Lambda}+\Omega
\ee
so that the meaning of each separate term in (\ref{eq18}) is not important, but what matters is their sum which should contribute about 73\% to the total balance in (\ref{eq17}).
Note also that as in the previous case of DE dominance, $\alpha$ and $\beta$ appear in the combination $(\alpha-1)/\beta$ in (\ref{eq15}), and hence we can define the parameter $\gamma=(\alpha-1)/\beta$. This effectively reduce the number of the free parameters of the model to three, as can be seen from (\ref{eq15}) by using the above replacement for $\alpha$ and $\beta$ and the flatness condition (\ref{eq17}). At the same time this simplifies the procedure of fitting the parameters with the current observational data.
In terms of the redshift $z$ the Hubble parameter takes the form 
\be\label{eq19}
\tilde{H}^2=\Omega'_{m0}(1+z)^3+\Omega_{\Lambda}+\Omega (1+z)^{2\gamma},
\ee
and the respective total EoS becomes
\be\label{eq20}
\text{w}_t(z)=-1+\frac{2\gamma\Omega(1+z)^{2\gamma}+3\Omega'_{m0}(1+z)^3}{3\left[\Omega(1+z)^{2\gamma}+\Omega'_{m0}(1+z)^3+1-\Omega'_{m0}-\Omega\right]}
\ee
where the flatness condition (\ref{eq17}) was used. 
Before beginning with the fixing of the parameters, we illustrate the behavior of the model by choosing some representative values of the parameters, guided by the redshift transition in the deceleration acceleration parameter and by the present observational value of the dark energy EoS parameter. 
From (\ref{eq19}) the  deceleration parameter is given by 
\be\label{eq21}
q(z)=-1+\frac{2\gamma\Omega(1+z)^{2\gamma}+3\Omega'_{m0}(1+z)^3}{2\left[\Omega(1+z)^{2\gamma}+\Omega'_{m0}(1+z)^3+1-\Omega'_{m0}-\Omega\right]}
\ee
The condition $q(z_T)=0$ gives $\Omega$ in terms of $\Omega'_{m0}$ and $\gamma$:
\be\label{eq22}
\Omega=\frac{2(1-\Omega'_{m0})-\Omega'_{m0}(1+z_T)^3}{1+(\gamma-1)(1+z_T)^{2\gamma}}
\ee
Let's turn to the equation of state of the dark energy sector.\\
From (\ref{eq19}) it follows the expression for the holographic dark energy density as
\be\label{eq24}
\rho_{DE}=3H_0^2\left(\Omega_{\Lambda}+\Omega (1+z)^{2\gamma}\right)
\ee
using the continuity equation $\dot{\rho_{DE}}+3H(\rho_{DE}+p_{DE})=0$ in terms of the redshift $z$ we find the DE pressure density, and the following expression for the EoS of the DE
\be\label{eq25}
\text{w}_{DE}=\frac{p_{DE}}{\rho_{DE}}=-1+\frac{(1+z)}{3}\frac{1}{\rho_{DE}}\frac{d\rho_{DE}}{dz}=-1+\frac{1}{3}\frac{2\gamma\Omega(1+z)^{2\gamma}}{\Omega_{\Lambda}+\Omega(1+z)^{2\gamma}}
\ee
at $z=0$ (\ref{eq25}) becomes
\be\label{eq26}
\text{w}_{DE_{0}}=-1+\frac{1}{3}\frac{2\gamma\Omega}{1-\Omega'_{m0}}
\ee
where the flatness condition (\ref{eq17}) was used. Assuming the current observational restrictions on the DE EoS \cite{hicken}, we can write $\text{w}_{DE_{0}}=-1+\delta$, where $\delta$ represents the deviation from $-1$ which may be in the range $-0.2<\delta<0.2$. Replacing in (\ref{eq26}) we obtain the relation
\be\label{eq27}
\delta=\frac{1}{3}\frac{2\gamma\Omega}{\Omega_{\Lambda}+\Omega}=\frac{1}{3}\frac{2\gamma\Omega}{1-\Omega'_{m0}}
\ee
for a given $\delta$ and $z_T$, and using the relations (\ref{eq17}), (\ref{eq22}) and (\ref{eq27}), the number of free parameters appearing in the solution (\ref{eq19}) reduces to one. Note that for $\delta=0$ it follows that $\gamma=0$ or $\Omega=0$, both of them conducing to the cosmological constant. And negative values of $\delta$ (indicating quintom behavior of the model), are possible if $\gamma<0$ (or even $\gamma>0, \Omega<0$). \\
Replacing $\Omega$ from (\ref{eq27}) in (\ref{eq22}) we obtain an equation to determine $\Omega'_{m0}$ for a given $z_T$, $\delta$ and $\gamma$:
\be\label{eq28}
3\delta(1-\Omega'_{m0})\left[1+(\gamma-1)(1+z_T)^{2\gamma}\right]=\gamma\left[2(1-\Omega'_{m0})-\Omega'_{m0}(1+z_T)^3\right]
\ee
Thus, solving the Eq. (\ref{eq28}) for  $\delta=0.05$, $zT=0.75$ and $\gamma=0.1$ gives $\Omega'_{m0}=0.273$. Quintom behavior can be obtained for $\delta=-0.05$ (i.e. $\text{w}_{0DE}=-1.05$), $zT=0.75$ and $\gamma=-0.1$, giving $\Omega'_{m0}=0.269$. 
From (\ref{eq27}) $\Omega$ takes the values $0.545$ and $0.548$ respectively, and the dark energy density parameter can be found from: $\Omega_{DE}=\Omega_{\Lambda}+\Omega=1-\Omega'_{m0}$, completing in this manner the determination of the original model parameters as displayed in (\ref{eq19}). To obtain $\Omega'_{m0}$ we have assumed concrete values for $\gamma$, but we will use the observational data to fix the three parameters of the model, without assuming any prior values. \\
%%%%%%%%%%%%%%%%%%%%%%%%%%%%%%%%%%%%%%%%%%%%%%%%%%%%%%%%%%%%%%%%%%%%%%%%%%%%%%%%%%%%%%%%%%%%%%%%%%%%%%%%%%%%%%%%%%%%%%%%%%%%%%%%%%%%%%%%%%%%%%%%%%%
{\bf{Cosmological constraints from SNIa, CMB and BAO.}}\\
We turn now to use the data sets to constrain the parameters $\Omega'_{m0}$, $\text{w}_{DE_{0}}$ and $\gamma$. We performed a $\chi^2$ analysis to obtain the constraints on $\Omega'_{m0}$ $\text{w}_{DE_{0}}$ and $\gamma$ using the Constitution (397) SNIa data set \cite{hicken}, the shift parameter of the cosmic microwave background (CMB) form WMAP5 observations \cite{komatsu} and the baryon acoustic oscillations from the Sloan digital sky survey (SDSS) \cite{einsenstein}. The details of the $\chi^2$ technique and definitions of the main observational quantities used to constraint the cosmological parameters, can be found in \cite{einsenstein,bond,wang,nesseris,nesseris1} (see \cite{li1} for holographic dark energy models). We will just quote here that the luminosity distance times $H_0$ for our model is given by 
\be\label{eq30}
d_L(z)=(1+z)\int_0^z\frac{cdz'}{\tilde{H}(z',\Omega'_{m0}, \text{w}_{DE_{0}}, \gamma)}
\ee
where $\tilde{H}(z,\Omega'_{m0}, \text{w}_{DE_{0}}, \gamma)$ is given by  (\ref{eq19}), after replacing $\Omega$ in terms of $\text{w}_{DE_{0}}$ through the relation (\ref{eq26}).
After minimizing the $\chi^2$ function with respect to the model parameters $\Omega'_{m0}$, $\text{w}_{DE_{0}}$ and $\gamma$, the following best fit values were found : $\Omega'_{m0}=0.272$, $\text{w}_{DE_{0}}=-0.962$, $\gamma=0.9$. This values define the best fit value for $h$: (the Hubble constant $H_0$ in units of 100 km/s/Mpc) $h=0.649$
The behavior of the Hubble parameter against the experimental bars  (see \cite{stern} and references therein)  is shown in Fig. 1.
\begin{center}
\includegraphics[scale=0.7]{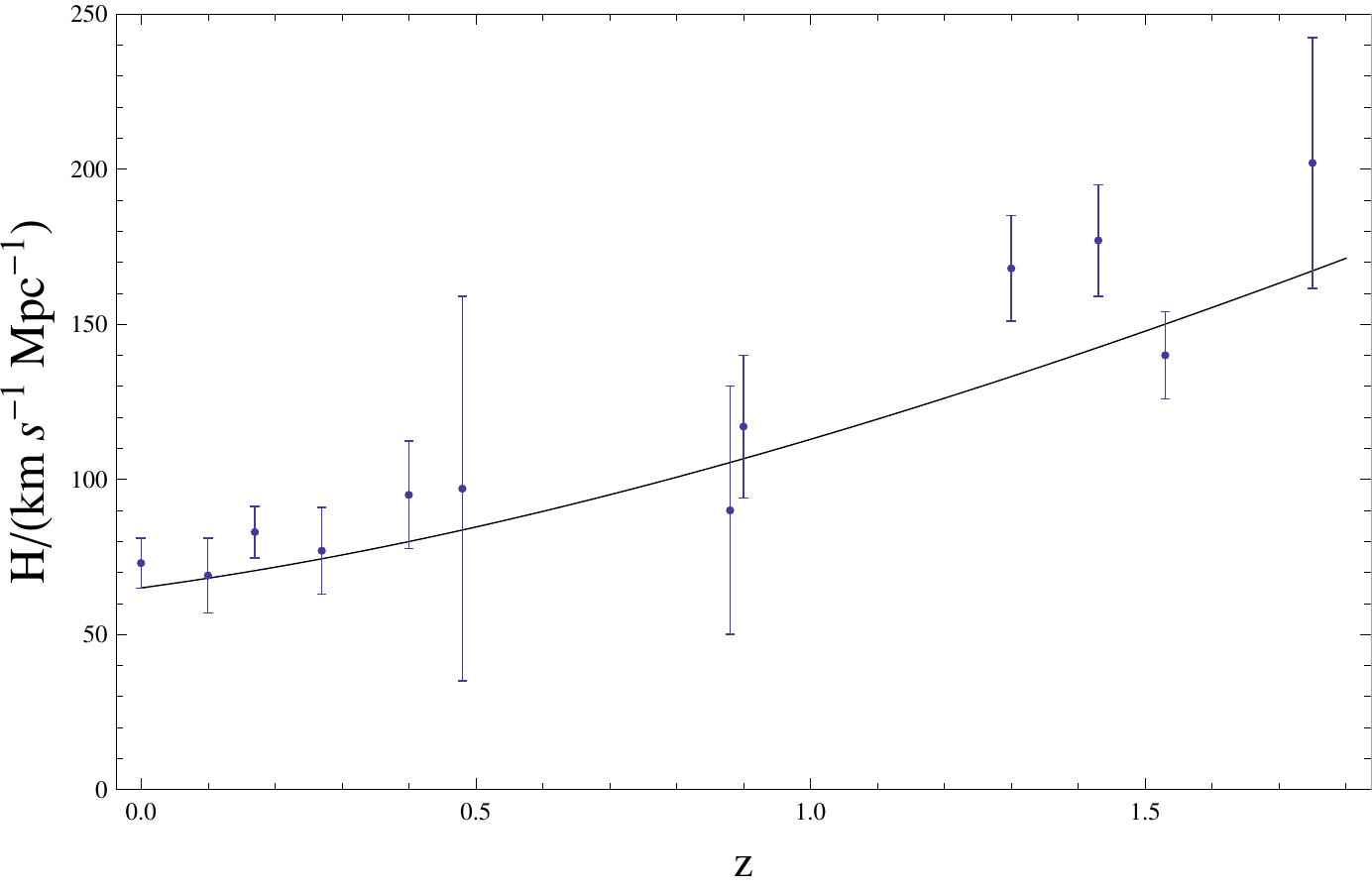}
\end{center}
\begin{center}
{Fig. 1 \it The behavior of $H$ for the best fit values $\Omega'_{m0}=0.272$, $\text{w}_{DE_{0}}=-0.962$ and $\gamma=0.9$, with $h=0.65$. The error bars correspond to the sample of passively evolving galaxies} \cite{stern}. 
\end{center}
%%%%%%%%%%%%%%%%%%%%%%%%%%%%%%%%%%%%%%%%%%%%%%%%%%%%%%%%%%%%%%%%%%%%%%%%%%%%%%%%%%%%%%%%%%%%%%%%%%%%%%%%%%%%%%%%%%%%%%%%

\section{discussion}
In this work we propose a new model of holographic density which incorporates a constant term as the first approximation to the infrared (IR) cut-off, and has notorious advantages with respect to the previous models, as it contains the $\Lambda$CDM model and additional DE term which could play an important role in the late time dynamics of the universe. 
The cosmological constant in the context of the holographic principle have been discussed in \cite{horava} (where using thermodynamical arguments it was showed that the most probable value of the cosmological constant is zero, but a very small cosmological constant is still allowed), and the presence of a constant term in the holographic density brings a quantum gravity nature to the  cosmological constant. Moreover, the current astrophysical data indicates that the constant cosmological constant adjusts very well to the observations, and this fact supports the presence of constant term in the holographic density. Adding this constant term to the IR cut-off may improve even models like the Hubble scale which was unable to produce accelerated expansion \cite{hsu}, \cite{li}. In this case if we add a constant term to the squared Hubble scale, we obtain the $\Lambda$CDM model. 
In order to reduce the number of free parameters, we grouped the terms proportional to $e^{-3x}$ in an effective matter density parameter, and the dependence on the initial model parameters $\alpha$ and $\beta$ reduces to dependence on the combination $\gamma=(\alpha-1)/\beta$. By using the astrophysical data we can fix $\gamma$, and this means that one of the parameters $\alpha$ or $\beta$ still free. If for instance we let $\beta$ free, then the relevant cosmological parameters which dictate the dynamical evolution of the universe do not depend on $\beta$, and there is apparently not dynamical criteria to fix $\beta$. 
To fix $\beta$ we can use geometrical criteria in order to convert the last two terms in (\ref{eq5}) in a term proportional to the Ricci scalar. In this case the constant $\beta$ is automatically defined. From the relation $\alpha=\gamma \beta+1$, if we set $\beta=1/(2-\gamma)$ then the last two terms in (\ref{eq5}) become $1/(2-\gamma)(2H^2+\dot{H})$.\\
We have also shown that the holographic density contributes to the matter density defining in this way the ``effective'' matter density containing the usual baryonic and dark matter (see (\ref{eq14}-\ref{eq16})). We can interpret this result as meaning that the present holographic density may account for the dark matter content, and in this sense previously considered models of holographic DE based on local quantities $H^2$ and $\dot{H}$ can give rise only to constant DE EoS parameter.
The model parameters ($\Omega'_{m0}, \Omega, \gamma$) were constrained using the observational data including the Constitution sample of SNIa, the CMB shift parameter given by WMAP5, and BAO measurement from SDSS. We found the best fit values  $\Omega'_{m0}= 0.272$, $\text{w}_{DE_{0}}=-0.962$ and $\gamma=0.9$, with $\chi^{2}_{min}=465.941$. The Hubble parameter versus the redshift is plotted in fig. $1$. A detailed study of the available data to constraint the parameters will be done elsewhere.\\
Resuming, the present model could be relevant because of its simplicity; it introduces a ``constant'' cosmological term in the infrared cut-off of the holographic density, which puts the cosmological constant in the context of the holographic principle, attributing quantum gravity nature to the cosmological constant; it improves previously considered holographic models of DE based on local quantities; it may reproduce the standard $\Lambda$CDM model even without explicitly giving matter content (see (\ref{eq8}) at $(\alpha-1)/\beta=3/2$).
We expect that future high precision experiments allow accurately determine the parameters of the model and define its quintessence or quintom nature.

\section*{Acknowledgments}
This work was partially supported by Fundacion SENECA (Spain) in the frame of PCTRM 2007-2010.


\begin{thebibliography}{99}
\bibitem{riess} A.G. Riess, et al., Astron. J. \textbf{116}, 1009 (1998), [astro-ph/9805201]
\bibitem{riess1} A.G. Riess, et al., Astron. J. \textbf{117},707 (1999)
\bibitem{perlmutter} S.Perlmutter et al, Nature \textbf{391}, 51 (1998)
\bibitem{perlmutter1} S. Perlmutter et al. [Supernova Cosmology Project Collaboration], Astrophys. J. \textbf{517}, 565 (1999) [astro-ph/9812133]
\bibitem{astier} P. Astier et al., Astron. Astrophys. \textbf{447}, 31 (2006) [astro-ph/0510447].
\bibitem{kowalski} M. Kowalski, et. al., Astrophys. Journal, \textbf{686}, p.749 (2008), arXiv:0804.4142
\bibitem{amanullah} R. M. Amanullah et. al. , arXiv:1004.1711 [astro-ph.CO]
\bibitem{hicken} M. Hicken et al., Astrophys. J. {\bf700}, 1097,2009; arXiv:0901.4804v3
\bibitem{spergel} D. N. Spergel et al. [WMAP Collaboration], Astrophys.
J. Suppl. 148, 175 (2003) [astro-ph/0302209]
\bibitem{spergel1} D. N. Spergel et al., Astrophys. J. Suppl. \textbf{170}, 377 (2007); astro-ph/0603449.
\bibitem{komatsu} E. Komatsu et al. [WMAP Collaboration], Astrophys. J. Suppl. \textbf{180}, 330 (2009), arXiv:0803.0547
\bibitem{larson} D. Larson et al., Astrophys. J. Supp. \textbf{192}, 16 (2011); arXiv:1001.4635v2 [astro-ph.CO]
\bibitem{tegmark} M. Tegmark et al. [SDSS Collaboration], Phys. Rev. D 69, 103501 (2004) [astro-ph/0310723]
\bibitem{tegmark1} M. Tegmark et al. [SDSS Collaboration], Astrophys. J. \textbf{606}, 702 (2004), astro-ph/0310725
\bibitem{abazajian} K. Abazajian et al. [SDSS Collaboration], Astron. J. \textbf{128}, 502 (2004) [astro-ph/0403325]
\bibitem{abazajian1} K. Abazajian et al. [SDSS Collaboration], Astron. J. \textbf{129}, 1755 (2005) [astro-ph/0410239].
\bibitem{tegmark2} U. Seljak et al. [SDSS Collaboration], Phys. Rev. D \textbf{71}, 103515 (2005), astro-ph/0407372.
\bibitem{einsenstein} D. J. Eisenstein et al. [SDSS Collaboration], Astrophys. J. 633, 560 (2005) [astro-ph/0501171].
\bibitem{tegmark3} M. Tegmark et al. [SDSS Collaboration], Phys. Rev. D \textbf{74}, 123507 (2006), astro-ph/0608632.
\bibitem{peebles} P.J.E. Peebles and B. Ratra, Rev. Mod. Phys. \textbf{75} (2003) 559 [astro-ph/0207347]
\bibitem{copeland} Edmund J. Copeland, M. Sami and Shinji Tsujikawa, Int. J. Mod. Phys. D \textbf{15}
1753-1936 (2006), arXiv:hep-th/0603057
\bibitem{turner} J.A. Frieman, M.S. Turner and D. Huterer, Ann. Rev .Astron. Astrophys. \textbf{46}, 385 (2008), arXiv:0803.0982[astro-ph]
\bibitem{sahni} Varun Sahni and Alexei Starobinsky, Int.J.Mod.Phys.D\textbf{9}, 373-444, 2000, astro-ph/9904398; 
\bibitem{sahnii} V. Sahni, 	Lect. Notes Phys. \textbf{653}, 141-180 (2004), arXiv:astro-ph/0403324v3
\bibitem{padmanabhan} T. Padmanabhan, Phys. Rept \textbf{380}, 235 (2003), [hep-th/0212290].
\bibitem{bekenstein} J. D. Bekenstein, Phys. Rev. D \textbf{7}, 2333 (1973).
\bibitem{thooft} G. 't Hooft, [gr-qc/9310026].
\bibitem{bousso} R. Bousso, \textit{JHEP} \textbf{9907}, 004 (1999), [hep-th/9905177].
\bibitem{cohen} A. Cohen, D. Kaplan and A. Nelson, Phys. Rev. Lett. \textbf{82}, 4971 (1999), [hep-th/9803132].
\bibitem{susskind} L. Susskind,  J. Math. Phys. (N. Y) \textbf{36}, 6377 (1994).
\bibitem{fischler} W. Fischler and L. Susskind, [hep-th/9806039].
\bibitem{hsu} S. D. H. Hsu, Phys. Lett. B \textbf{594}, 13 (2004), [hep-th/0403052].
\bibitem{li} M. Li, Phys. Lett. B \textbf{603}, 1 (2004), [hep-th/0403127].
\bibitem{sergei} E. Elizalde, S. Nojiri, S.D. Odintsov and P. Wang, Phys. Rev. D \textbf{71}, 103504 (2005), hep-th/0502082
\bibitem{sergei1} S. Nojiri and S. D. Odintsov, Gen. Rel. Grav. \textbf{38}, 1285 (2006), hep-th/0506212.
\bibitem{granda} L. N. Granda and A. Oliveros, Phys. Lett. B \textbf{669}, 275 (2008), gr-qc/0810.3149. 
\bibitem{granda2} L. N. Granda and A. Oliveros, Phys. Lett. B \textbf{671}, 199 (2009), gr-qc/0810.3663. 
\bibitem{granda3} L. N. Granda, Int.J.Mod.Phys.D, V.18, No.11, 1749 (2009); arXiv:0811.4103 [gr-qc]
\bibitem{granda4} L. N. Granda and A. Oliveros, arXiv:0901.0561 [hep-th]
\bibitem{horava} P. Horava and D. Minic, Phys. Rev. Lett. \textbf{85}, 1610 (2000), hep-th/0001145
\bibitem{bond} J. R. Bond, G. Efstathiou and M. Tegmark, Mon. Not. Roy. Astron. Soc. \textbf{291}, L33 (1997), astro-ph/9702100
\bibitem{wang} Y. Wang and P. Mukherjee, Astrophys. J. \textbf{650}, 1 (2006), astro-ph/0604051
\bibitem{nesseris} S. Nesseris and L. Perivolaropoulos, Phys.Rev.D\textbf{72}, 123519 (2005); arXiv:astro-ph/0511040
\bibitem{nesseris1} S. Nesseris and L. Perivolaropoulos, JCAP 0702, (2007) 025 ; arXiv:astro-ph/0612653
%\bibitem{granda5} L. N. Granda, W. Cardona and A. Oliveros, arXiv:0910.0778 [hep-th]
%\bibitem{zhang} X. Zhang, arXiv:0901.2262 [astro-ph], Phys. Rev. D\textbf{79}, 103509 (2009)
\bibitem{li1} M. Li, X. Li, S. Wang and X. Zhang, JCAP 0906, (2009) 036; arXiv:0904.0928
\bibitem{stern} D. Stern et al., JCAP 02 (2010) 008
%\bibitem{nesseris} S. Nesseris and L. Perivolaropoulos, JCAP. 0702, 025 (2007).
%Elisa Di Pietro, Jean-Francois Claeskens ; Mon.Not.Roy.Astron.Soc. 341 (2003) 1299; arXiv:astro-ph/0207332v3

\end{thebibliography}
\end{document}